# A New GAUSS Software Component for Implementing Dynamic Tests of Contagion with an Application

**Abdulnasser Hatemi-J[1] and Alan Mustafa[2]**

1Department of Economics and Finance, College of Business and Economics, UAE University, Al Ain P.O. Box 15551, United Arab Emirates. Email: AHatemi@uaeu.ac.ae

2 Da Vinci Institute, Duhok, Kurdistan Region, Iraq. Email: Alan.Mustafa@ieee.org

**Abstract:** The main goal of this article is to present a new software component created by the authors in GAUSS for implementing dynamic tests for contagion. These tests are aimed at providing the possibilities of testing for spiller-over effects between financial markets during a crisis or during a pandemic situation such as the one that we experienced recently. Unlike standard tests for contagion that have fixed dates for a potential shift, the tests implemented here are dynamic in the sense that the potential contagion effect is measured across the time span. The leverage bootstrap simulation technique is utilized in order to generate critical values that are reliable when the tests are implemented in order to make sure that the conducted inference is accurate. Since the approach is new, software component needs to be created for conducting the underlying tests, which is accomplished via this article. The module is very easy to implement by the general practitioners via a graphical user interface. An application is offered by using real data from stock markets in order to demonstrate how the software component can be used.

**Keywords**: Software development, Financial Crisis, Stock markets, Leverage bootstrap, Time series, Gauss.

**JEL Classifications**: C22, C61, G17

## 1. Introduction

The relationship between random financial variables is of predominant importance especially crises periods. Investors, financial institutions and policymakers are enormously interested in discovering whether there are spill-over or contagion impacts from one given market to toother markets during a period of distress. The potential contagion effect across borders is increasingly gaining attention in a time that is characterized by increasing financial markets integration across the globe in order tom curtail the potential negative spill-over effects. The financial markets are depicting higher volatility than ever before. However, there is no unanimously acknowledged concept for describing contagion as a well-defined phenomenon in the existing literature. According to Forbes and Rigobon (2002) [1] a popular definition of contagion that has historically been utilized is an increase in the relationship between markets when a sudden shock takes place in a given market, which is statistically significant. For some applications of this approach see inter alia: King and Wadhwani, 1990 [2]; Lee and Kim, 1993 [3]; Calvo and Reinhart, 1996 [4]; Forbes and Rigobon, 2002 [1], Hon et al., 2004 [5]; Pretorius and Beer, 2004 [6].[1]

The main goal the paper is to present a software component created in Gauss for implementing the dynamic tests for contagion. These tests are aimed at providing the possibilities of testing for spiller-over effects between financial markets during a crisis or during a pandemic situation such as the one that we experienced. The tests are dynamic in the sense that the potential contagion effect is measured across the time span. The leverage bootstrap simulation technique is utilized in order to generate critical values that are reliable when the tests are implemented in order to make sure that the conducted inference is accurate. Since the approach is new, the new software is needed for implementing the underlying tests.

## 2. Methodology

In order to implement tests for the contagion effect of variable X on variable Z, the following regression method is used:

$$Z_t = a_1 + a_2 D_t + b_1 X_t + b_2 D_t X_t + e_t \tag{1}$$

Where $a_1$ is the intercept, $a_2$ represents the intercept, Dt is an indicator variable that undertakes the value equal to zero and it assume value one for the period of after the break. The error term is denoted by $e_t$. Equation (1) can be written compactly by using the matrix algebra as the following:

1 For additional pertinent publications on the topic see among others Boschi (2005) [7], Boyer, Gibson and Loretan (1999) [8], Collins and Gavron (2004) [9], Davison and Hinkley (1999) [10], Eichengreen, Rose and Wyplosz (1996) [11], Granger, Huang and Yang (2000) [12] and Hamao, Masulis and Ng (1990) [13], Forbes, K. J. (2004) [14], Longin, F. M. and Solnik, B. (1995) [15], Loretan, M. and English, W. B. (2000) [16], Yamori, N. (1999) [17].

$$Y = CW + e \quad (2)$$

Where

$$Y = \begin{bmatrix} Z_1 \\ Z_2 \\ \vdots \\ Z_T \end{bmatrix}, W = \begin{bmatrix} 1 & D_1 & X_1 & D_1X_1 \\ 1 & D_2 & X_2 & D_2X_2 \\ \vdots & \vdots & \vdots & \vdots \\ 1 & D_T & X_T & D_TX_T \end{bmatrix}, C = [a_1 \quad a_2 \quad b_1 \quad b_2] \text{ and } e = \begin{bmatrix} e_1 \\ e_2 \\ \vdots \\ e_T \end{bmatrix}$$

The parameter vector *C* can be estimated by using the ordinary least squares (OLS) approach as the following:

$$\hat{C} = (W'W)^{-1}W'Y \quad (3)$$

The pairwise bootstrap technique is implemented according to the following steps:

1. Start with generating the bootstrap values for both the dependent and the independent variables (denoted by $Y^*$ and $W^*$) via resampling with replacements.
2. Next, Estimate the parameter vector C using the values by $Y^*$ and $W^*$ as

$$\hat{C}^* = (W^{*\prime}W^*)^{-1}W^{*\prime}Y^* \quad (4)$$

3. Third, repeat steps one and two *N* times.
4. Fourth, obtain the casewise bootstrap parameter vector $\left(\overline{\hat{C}^*}\right)$ by taking the average value of all *N* estimators.
5. Lastly, redo steps one to four by using sub-samples in order the make the tests dynamic. We have chosen to the start from the sub-sample that covers 45% of the entire period. However, the software component is flexible regarding this issue. The practitioner can choose any other percentage that might be preferred.

The bootstrap approach is also used to test the significance tests. The bootstrap p-values are estimated. In order to consider the potential prospect of the change in the relationship between the underlying variables, tests need to be implemented by utilizing subsamples. In order to estimate the parameters efficiently and conduct significance tests accurately in the presence of non-normality and time-varying volatility, which is the case during a financial crisis, Hatemi-J and Hacker (2005) [18] suggest using the pairwise leverage bootstrap simulations. This technique does not assume any a priory distribution for the underlying data and it makes used of the empirical distribution, which is more accurate than the asymptotics ones. The simulation results show that this test method has correct size properties and higher power compared to a test that is based on an asymptotic distribution. In this original version of the test is it required that the period of the break is know in advance. However, since this issue might be too restrictive and the exam timing of the break might not be known exactly in advance, Hatemi-J (2025) [19] suggests making the contagion test dynamic in such a way that the potential break is determined endogenously. By this way it is also possible to measure the potential contagion effect across time and find out empirically when the spill-over effect is the highest. Thus, this approach is more informative.

## 3. Description of the Module

The software component that can be used for implementing the dynamic contagion tests using the pairwise bootstrap simulations is created by Mustafa and Hatemi-J (2025) [20] in Gauss, which is available online. In this section we present our software component in order to demonstrate how it can be used. An application using real data in order to capture the impact of oil prices on the world price index on a daily basis is provided.

### *3.1 Data Flow Chart*

The process as indicated in the Figure 1 starts by loading a dataset and creating two subsets of arrays. The first set includes a smaller portion of the dataset (in this application 45% of the whole dataset were selected). To run and evaluate the test for calculated coefficients for each range of dataset, a loop was designed to iterate the process for different ranges. After calculating the OLS (the ordinary least squares) and sorting the coefficients, their means and medians can be calculated using the pairwise bootstrap simulation method. This method is also used for testing the significance of each estimated parameter. The p-values are presented. A sample of results of those calculations is presented in Figure 2 and Figure 3.

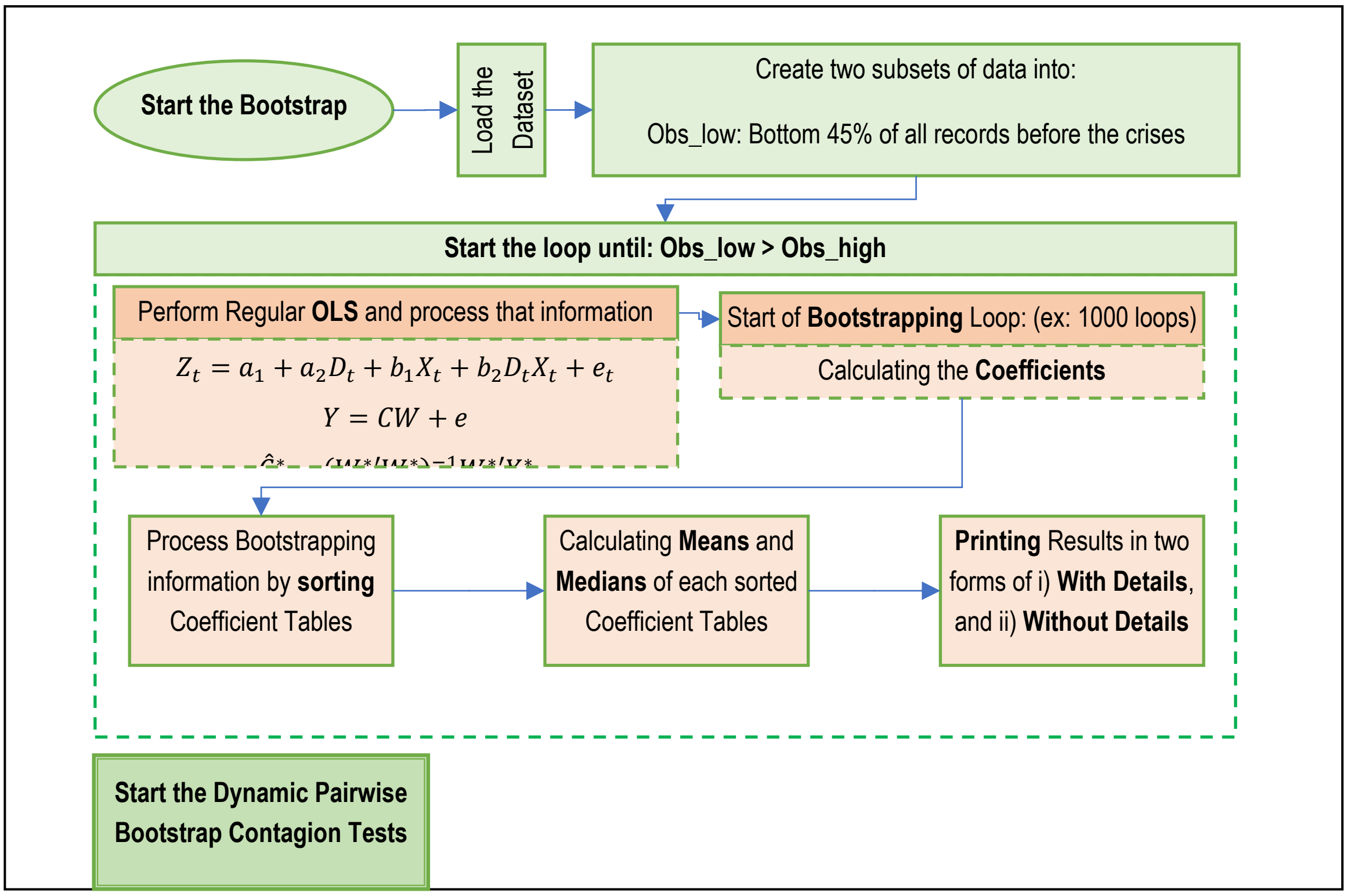

Figure 1: The Schematic diagram of dynamic pairwise bootstrap contagion tests. The code is published under the label GMDLBCT1.PRG.

### *3.2 Pseudocode*

The pseudocode for the schematics shown in Figure 1 is presented in below named Algorithm 1. This structure of this code contains three main sections of i) loading and creating datasets, ii) calculation of required data to create new information, and iii) printing results in two formats with and without details.

**Algorithm I:** The Pseudocode for GMDLBCT

```
Load Dataset
Create two subsets of data:
Declaring and Setting Initial and Ending Samples;
obslow = buttom 45% of all records;
obshighO = 55% of all records;
theRange = obshighO - obslow + 1;

Start the loop until: Obs_low > Obs_high
Do until obslow > obshighO;
        obshigh = obstot - obslow;
Y= w[.,1];
        X =w[.,2];
        D = zeros(obslow,1)|ones(obshigh,1);

        Perform Regular OLS and process that information
        Z = ones(obstot,1)~D~X~(X.*D);
coeff = zeros(rmax,cols(Z));
InvZTZ= Inv(Z'Z);
        Ahat = InvZTZ*(Z'*Y);
varres = Y - Z*Ahat;
t = Ahat./(sqrt((varres'varres/(obstot-cols(Z)))*diag(InvZTZ)));
```

adjres =varres;

Coeff_n = 0; (where 1<= n <=3)
***Start of Boostrapping Loop: r counts it***
Calculating the Coefficients
Do until r > rmax;
randomnumbers = rndu(1,obstot);
index = 1+ trunc(obstot*randomnumbers);
Yboot = Y[index,1]; Xboot = X[index,1]; Dboot = D[index,1];
Zboot = ones(obstot,1)~Dboot~Xboot~(Xboot.*Dboot);
Ahatboot = (Inv(Zboot'Zboot)*(Zboot'Yboot))';
Coeff_n = Coeff_n + Ahatboot[1, n];
Coeff[r,n] = Ahatboot[1,n];
r = r + 1;

***Process Bootstrapping Information (Sorting)***
Sortedcoeff_n = Sortc(Coeff[.,n],1);

***Mean Bootstrap Coefficients***
meansortedcoeff = meanc(sortedcoeff4);
medianc_n = median(sortedcoeff_n);
mnBC1[i,n] = Coeff_n/rmax; mdnBC1[i,n] = medianc_n;

***P-value Based on Empirical Distribution***
bpv1[i,n] related to the value of Medianc_n and the outcome comes from dependability value of (rmax, Sortedcoeff_n and medianc_n);

obslow = obslow + 1;
i = i + 1;
**Printing Results;**

### *3.3 Printing the Outcome*

Towards the end of calculations, the user has two options of printing the details for each step of calculation (Figure 2) or just printing the last results (Figure 3).

```
*******************************************************************************************************************

obslow =      2083.0000
obshigh =      1705.0000

OLS estimates=
     79.633244
    -94.103950
  0.00029750783
   0.059434495
************************************************************
************   Mean Bootstrap coefficients    **********
mean Bootstrap coefficient 1 =      79.672277        median Bootstrap coefficient 1=      79.638095
mean Bootstrap coefficient 2 =     -94.135775        median Bootstrap coefficient 2=     -94.131539
mean Bootstrap coefficient 3 =   0.00025794908       median Bootstrap coefficient 3 =  0.00025409683
mean Bootstrap coefficient 4 =     0.059462928       median Bootstrap coefficient 4 =    0.059487720
************************************************************
**********   P-value based on empirical distribution    **********
 Bootstrap p-value=     0.00000000
Bootstrap p-value==     0.00000000
Bootstrap p-value==     0.94830000
Bootstrap p-value==     0.00000000
*******************************************************************************************************************
```

Figure 2: printed results while calculating necessary information such as means and medians.

**Pairwise Bootstrap Test for Contagion**

| | FOR THE SLOPE | | | FOR THE CHANGE IN THE SLOPE | | |
|---|---|---|---|---|---|---|
| Sub-sample | Bootstrap P-Value | Mean Bootstrap Coefficient | Median Bootstrap Coefficient | Bootstrap P-Value | Mean Bootstrap Coefficient | Median Bootstrap Coefficient |
| 1.000 | 0.000 | −0.026 | −0.026 | 0.000 | 0.083 | 0.083 |
| 2.000 | 0.000 | −0.025 | −0.025 | 0.000 | 0.083 | 0.083 |
| 3.000 | 0.000 | −0.025 | −0.025 | 0.000 | 0.082 | 0.082 |
| 4.000 | 0.000 | −0.024 | −0.024 | 0.000 | 0.082 | 0.082 |
| 5.000 | 0.000 | −0.024 | −0.024 | 0.000 | 0.082 | 0.082 |
| 376.000 | 0.000 | 0.059 | 0.059 | 0.874 | 0.001 | 0.001 |
| 377.000 | 0.000 | 0.059 | 0.059 | 0.902 | 0.001 | 0.001 |
| 378.000 | 0.000 | 0.059 | 0.059 | 0.923 | 0.000 | 0.000 |
| 379.000 | 0.000 | 0.059 | 0.059 | 0.948 | 0.000 | 0.000 |

Figure 3: A sample of printout for the contagion pairwise bootstrap test.

## 4. Application

To test the application the dataset of daily values of the world stock price index and the oil prices for the period of November 9th 2018 until May 14th 2020 from Bloomberg database were selected.

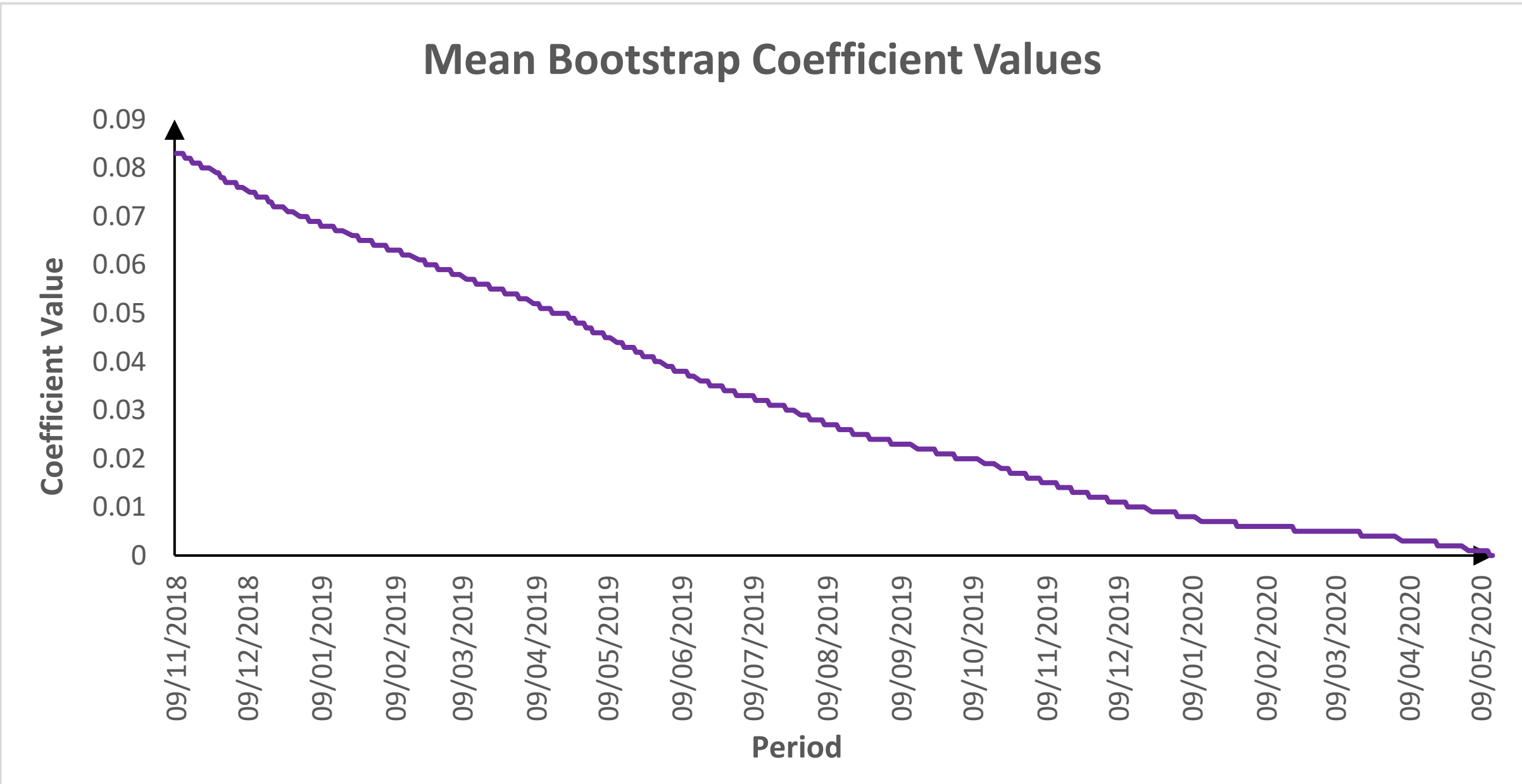

Graph 1: The estimated change in the slope parameter across time.

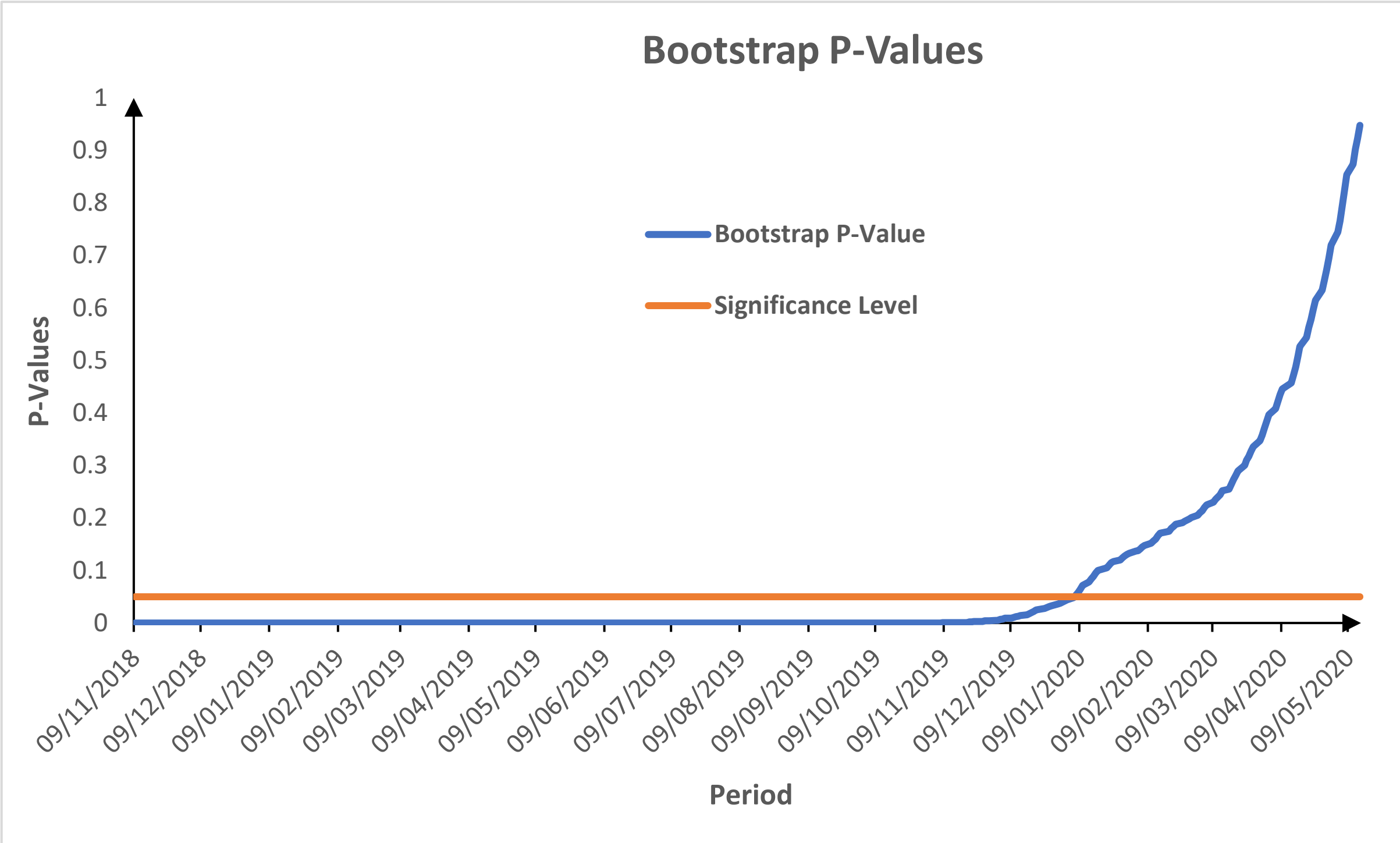

Graph 2: The estimated p-values of the null hypothesis of no change in the slope parameter across time. 5% is used as the significance level.

## 5. Conclusions

The need for implementing risk management strategies successfully increases enormously during the high impact and low probability events such as during a financial crisis. For accurate design of such hedging strategies and appropriate implantation of policies, measuring the interaction between the underlying financial variables or markets is essential. For this purpose, conducting contagion tests in order to capture the potential spill-over effects is of paramount importance. In this article we present a module created in Gauss that conducts dynamic contagion tests using pairwise bootstrap simulations. This approach has good size and power properties since it does not rely on any a priory distribution. An application is provided in order to investigate whether that impact of oil prices on the world stock prices shifts across time. The results show that there is indeed a significant change in the impact of oil prices on the world stock prices index until 9th of January 2020, which approximately represents the end of the COVID pandemic. The change in the slope parameter is positive across time, which indicates that the impact of the oil market on the stock market is increasing across time. The software component is very consumer friendly and it is available online.